\documentclass[numsec,webpdf,contemporary,large]{oup-authoring-template}%
\usepackage{lineno}
\begin{document}

\journaltitle{Preprint}
\DOI{DOI HERE}
\copyrightyear{2023}
\pubyear{year}
%\access{Advance Access Publication Date: Day Month Year}
\appnotes{Preprint}

\firstpage{1}

\title[Lost options commitment]{Lost options commitment: how short-term policies affect long-term scope of action}

\author[1,$\ast$]{Marina Mart\'inez Montero}%\email{marina.martinez@uclouvain.be}
\author[2]{Nuria Brede}%\email{nuria.brede@uni-potsdam.de}
\author[1]{Victor Couplet}%\email{victor.couplet@uclouvain.be}
\author[1]{Michel Crucifix}%\email{michel.crucifix@uclouvain.be}
\author[3,4]{Nicola Botta}%\email{botta@pik-potsdam.de}
\author[5]{Claudia Wieners}%\email{c.e.wieners@uu.nl}

\corresp[$\ast$]{Corresponding author. \href{email:marina.martinez@uclouvain.be}{marina.martinez@uclouvain.be}}

\address[1]{Earth and Life Institute, UCLouvain, Louvain-la-Neuve, Belgium}
\address[2]{University of Potsdam, Potsdam, Germany}
\address[3]{Potsdam Institute for Climate Impact Research, Potsdam, Germany}
\address[4]{Chalmers University of Technology, Göteborg, Sweden}
\address[5]{Institute for Marine and Atmospheric Research, Utrecht University, Utrecht, Netherlands}

%\received{Date}{0}{Year}
%\revised{Date}{0}{Year}
%\accepted{Date}{0}{Year}

%\editor{Associate Editor: Name}

\abstract{We propose to explore the sustainability of climate policies based on a novel commitment metric. This metric allows to quantify how future generations’ scope of action is affected by short-term climate policy. In an example application, we show that following a moderate emission scenario like SSP2-4.5 could commit future generations to heavily rely on carbon dioxide removal or/and solar radiation modification to avoid unmanageable sea level rise.}

\maketitle
\section{Main Text}\label{sec1}
Climate policy in the coming decades will have profound long-term impacts on global climate, ecosystems and human societies \cite{Clark2016, IPCCAR6-synthesis}. In this context, \emph{sustainability} is a critical consideration:
It pertains to meeting humanities' present needs without compromising the ability of future generations to meet their own. Thus, taking sustainability into account in policy assessments is essential to address climate change while also ensuring intergenerational fairness.

Given the long residence time of CO$_2$ in the atmosphere and its long-term impacts, taking fair decisions in the upcoming decades requires considering possible scenarios of anthropogenic forcing over timescales of centuries to millennia. 

However, in climate policy assessments based on integrated assessment models, longer timescales are often absent, or the long-term impacts of current decisions are heavily discounted relative to short-term impacts~\cite{Stern2007}.
Given that it is challenging to consider longer timescales for multiple reasons, this is understandable, but nevertheless unsatisfactory (or even ethically problematic~\cite{beckerman2007ethics, Portney2013}).

We therefore propose an approach to improving sustainability considerations in climate policy assessments. Our suggestion is to explore the sustainability of short-term climate policies based on a metric, which we call ``lost options commitment'', that quantifies future generations' scope of action to avoid harmful long-term futures. More particularly, we ask

\begin{quote}
	``Given a climate state that can be reached under realistic short-time emission scenarios, which long-term climate mitigation options are left to future generations to meet a specified climate target?''
\end{quote}

Our method is related to several well-established notions~\cite{ipcc_glossary} from the climate science literature: \emph{climate change commitment}~\cite{Wigley2005}, \emph{storylines}~\cite{shepherd2018storylines} and \emph{vulnerability}~\cite{wolf2013clarifying,Ionescu2016}.

Classical commitment assessments seek to quantify ``unavoidable'' climate impacts due to inertia in the climate system: a state of the climate system (typically the current one) may be called \emph{committed} to some future impact (like the amount of global warming or sea-level rise) under a given scenario~\cite{Wigley2005,Levermann2013,MacDougall2020,VanBreedam2020}. Lost options commitment focuses on the scope of action rather than the impacts: in a given state, humanity is \emph{committed by the lost options} to a narrower scope of action for meeting an intended climate target.

Commitment studies typically rely on few and simplified long-term scenarios, such as \emph{zero emissions}, \emph{constant composition} and \emph{constant emissions} \cite{tipes_ontology_2022,Wigley2005,Levermann2013,Clark2016,MacDougall2020,VanBreedam2020}. Such simplistic scenarios poorly capture human agency in reacting to climate change, which is one of the key aspects our metric attempts to capture. 
We adopt Shepherd et al.'s storyline approach~\cite{shepherd2018storylines} to compose representative sets of scenarios by combining a small number of building blocks. This modular approach enables the generation of a rich set of long-term scenarios.
For each individual long-term scenario, we may assess the commitment of a climate state with respect to a particular climate variable (like e.g.\ sea-level rise). However, as opposed to traditional commitment studies, we do not stop there. Instead, for a given state, we assess the compatibility of each long-term scenario with a chosen \emph{climate target} within a specified \emph{time horizon}. The climate target makes explicit a goal of decision making, like e.g., the 1.5$^\circ$C target of the Paris agreement. The time horizon can be thought of as the ``ethical time horizon'' of \cite{Lenton2008}.
Scenarios that are compatible with the climate target are considered as \emph{available options} in this state, the others as \emph{lost options}. 
We may say that in a given state, the \emph{loss of options commits} humanity to a narrower scope of action to meet the desired target. Therefore, we call this metric \emph{lost options commitment}, and call the percentage of lost options in a state its \emph{commitment level}.  

Lost options commitment is a generic metric in the sense that it can be instantiated with different combinations of long-term scenarios, climate target and timescale. However, it is also itself an instance of the general scheme for vulnerability metrics~\cite{wolf2013clarifying, Ionescu2016}.

We now propose that the sustainability of short-term policies (e.g., given in terms of SSP or RCP emission scenarios) can be analysed by measuring the commitment level of the states encountered when evolving an underlying climate model according to these policies.
We argue that by having more climate mitigation options at hand to avoid harmful outcomes (e.g., transgressing a specified target) humanity is in a better position for meeting its needs. Therefore, short-term policies that lead to loss of options, i.e., an increase in the commitment level, should be considered as \emph{unsustainable}.
On the other hand, if along the evolution corresponding to a short-term policy the commitment level decreases, this policy increases future generations' scope of action and can be considered as \emph{sustainable}. This approach is well-aligned with the IPCC's 'Window of opportunity to enable climate resilient development' illustrated in Figure~4.2 of the IPCC AR6 Synthesis Report~\cite{IPCCAR6-synthesis} and related to the concept of \emph{robustness} of decisions~\cite{WongRosenhead}.

We demonstrate our approach with a simple yet instructive example.
Our study uses as climate target \emph{avoiding sea level rise above 3m} within a time horizon of \emph{2000 years} and a set of 45 long-term scenarios as possible options. The horizon of the same magnitude as the ethical horizon suggested in \cite{Lenton2008}. The long-term scenarios are generated as combinations of 3 technologies together with usage variants:
\begin{itemize}
	\item \emph{Dec} -- Decarbonisation with rates \emph{slow}, \emph{medium}, \emph{fast}
	\item \emph{CDR} -- Carbon Dioxide Removal with intensities \emph{no}, \emph{weak}, \emph{strong}
	\item \emph{SRM} -- Solar Radiation Modification with possible intensities/durations \emph{no}, \emph{weak short}, \emph{strong short}, \emph{weak long}, \emph{strong long}
\end{itemize} 
For the concrete semantics of the scenario building blocks, see \ref{sec:methods}.
Given these ingredients, and an appropriate climate model, the lost options commitment can be computed for Earth system states containing a climate state and a current CO$_2$ emission rate. 
Here we focus on states that are reached along the SSP scenarios SSP1-2.6, SSP2-4.5, SSP3-7.0 and SSP5-8.5 from 2020 until 2100, every 10 years, see Fig.~\ref{fig:all_states}. Computing the commitment levels requires evolving each of these states along the 45 long-term scenarios for 2000 years, and checking which long-term scenarios comply with the climate target of avoiding sea-level rise above 3m. For this aim we use the reduced complexity model SURFER \cite{martinez_montero_surfer_2022}. This model has been designed for policy assessments with a long-term perspective, and links CO$_2$ emissions and solar radiation modification to climate change, ocean acidification and sea level rise. Other greenhouse gasses are absent in the presented example. 

\begin{figure}[t!]
	\centering
	\includegraphics[width=\linewidth]{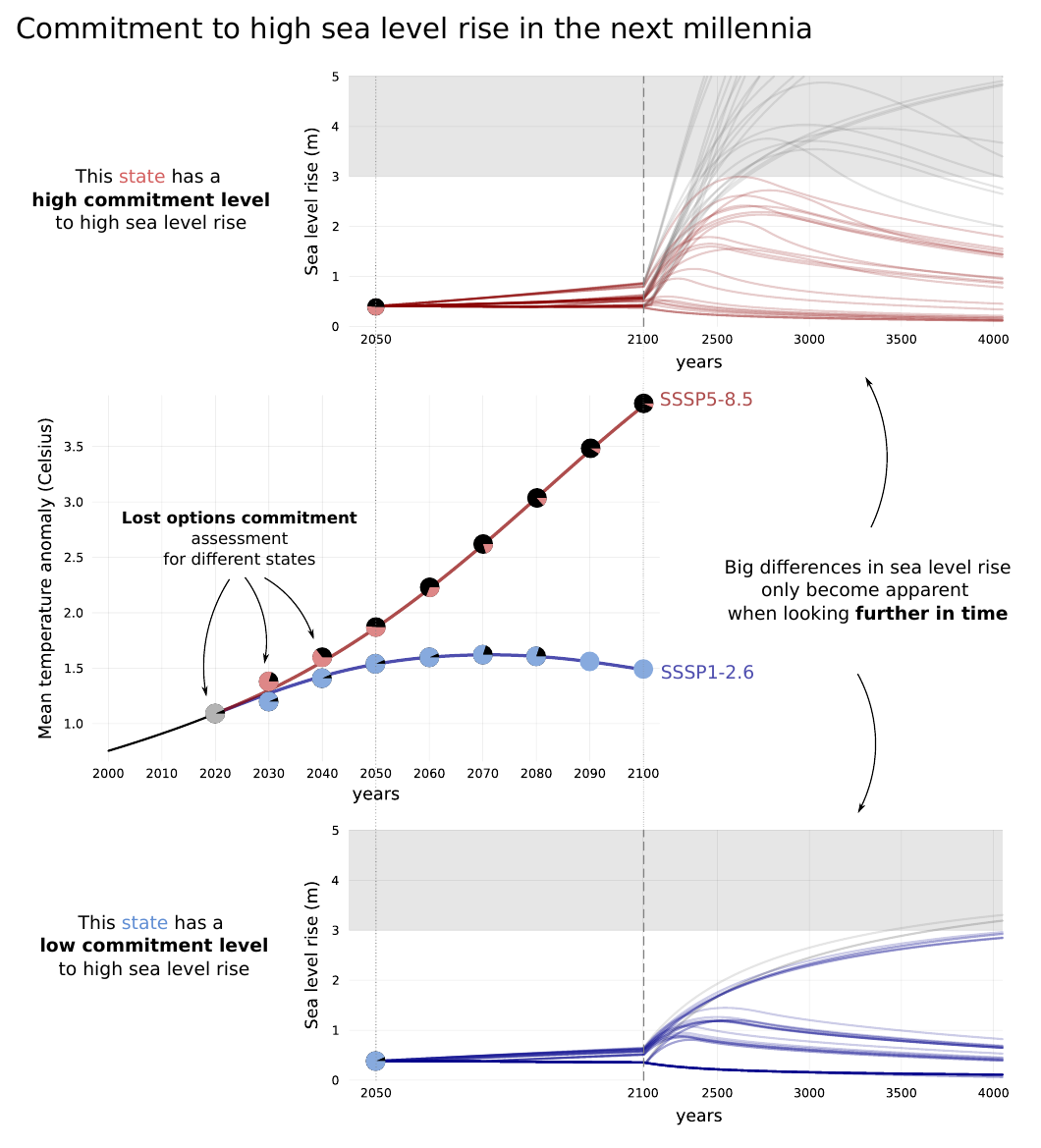}
	\caption{Lost options commitment with respect to high sea level rise in the next millennia. Assessment done for states encountered along SSP5-8.5 and SSP1-2.6. Central plot shows global mean temperature anomaly until 2100 along these scenarios. Pie-chart markers represent the commitment level of the different states: black corresponds to the fraction of long-term scenarios leading to high sea level rise and coloured corresponds to the fraction of long-term scenarios complying with the climate target. Upper and bottom right plots show the sea level rise trajectories for all long-term scenarios considered in the assessment for the states at year 2050 in SSP5-8.5 and SSP1-2.6 respectively. There we show the details of the metric used in this example: sea level rise above 3 meters with respect to pre-industrial anytime in the next 2000 years. Upper and bottom plots contain 45 curves, each of which corresponds to one of the considered possible long-term scenarios. Evolutions leading to high sea level rise are shown in grey and those complying with the climate target have been coloured according to the corresponding SSP scenario the state belongs to.}
	\label{fig:Fig1}
\end{figure}

Figure~\ref{fig:Fig1} shows that following SSP5-8.5 until 2050 leaves the Earth in a state in which more than half of the considered long-term scenarios fail to comply with the chosen climate target. Conversely, following SSP1-2.6 until 2050 results in a state with many available options. 
Note that sea level simulations up to 2100 do not capture the large differences expected for different states and long-term scenarios, hence the importance (for the chosen target) of looking further in time. 
States with similar temperature may have different commitment levels, e.g. SSP1-2.6 at 2040 and SSP5-8.5 at 2030 or SSP1-2.6 at 2050 and 2090. One reason for this are the different emission rates associated with these states. This highlights why climate policy negotiations and agreements should look beyond temperature targets.

\begin{figure}[t!]
	\centering
	\includegraphics[width=\linewidth]{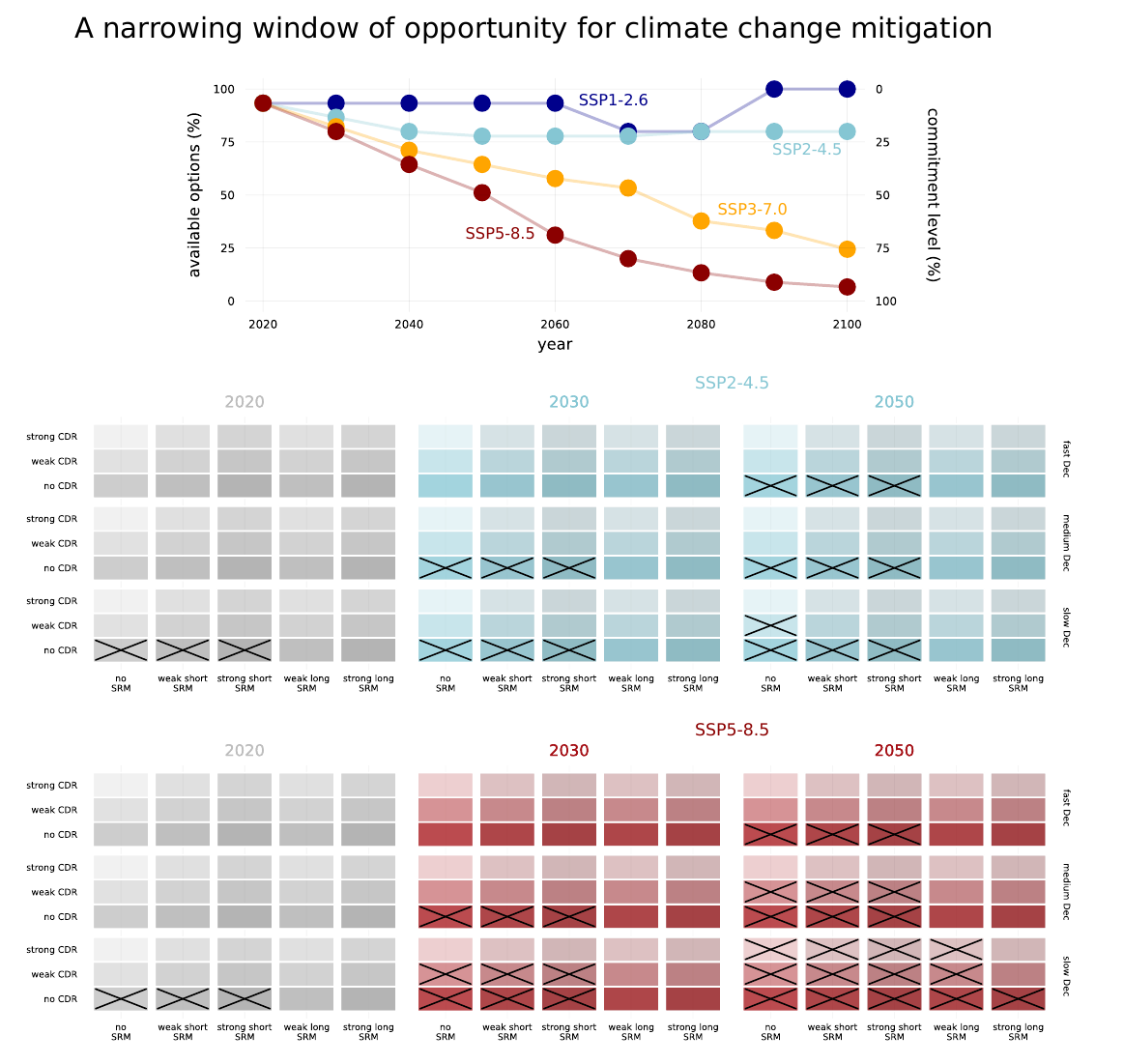}
	\caption{A narrowing window of opportunity for climate change mitigation. Top plot summarises the lost options commitment assessment for states along different SSP scenarios. The bottom plots show, for the states at 2020, 2030 and 2050 along SSP2-4.5 and SSP5-8.5, the outcomes of all the considered long-term scenarios: crossed-out options lead to high sea level rise for that particular state.
		Similar plots for all the states considered in this assessment can be found in Figs.~\ref{fig:SSP1}, \ref{fig:SSP2}, \ref{fig:SSP3} and \ref{fig:SSP5}.}
	\label{fig:Fig2}
\end{figure}

Figure~\ref{fig:Fig2} shows that, as expected, higher emission scenarios lead to fewer available options. The space of options plots for states along SSP2-4.5 and SSP5-8.5 in Fig.\ref{fig:Fig2}, show that in 2020, the \emph{slow Dec} options with \emph{no CDR} and with \emph{no SRM} or with \emph{short SRM} do not meet the chosen climate target. By 2030, this is also the case for options with \emph{medium Dec} and \emph{no CDR} with \emph{no SRM} or with \emph{short SRM}. This means that, starting a green transition by 2030, at a rate in line with the SSP scenarios (as the \emph{medium Dec}), will commit future generations to the use of \emph{CDR} or \emph{long SRM} to avoid long-term high sea level rise. By 2050 the use of \emph{CDR} or \emph{long SRM} is inescapable for states reached through SSP2-4.5 and SSP5-8.5 if the climate target is to be respected. Even if most options remain available for SSP2-4.5 at 2050, following this path until 2050 commits future generations to using \emph{CDR} or \emph{long SRM}. Notice that along scenario SSP1-2.6 (top plot in Fig.~\ref{fig:Fig2}), the commitment level decreases towards the end of the century, with all options becoming available by 2090. This happens because SSP1-2.6 includes CDR from $\sim$2075, see Fig.~\ref{fig:all_states}. 
For questions concerning sensitivity of the results to the specific construction, see the Supplementary Information Sec.~\ref{sec:suplementary}. 

Despite the simplicity of the example, we think that our study provides valuable insights into the sustainability of the considered SSP scenarios. Thanks to its systematic construction based on an explicit climate target, time horizon and set of scenarios, the lost option commitment metric is transparent and modular. It also has the advantage of informing policymakers about available options without being prescriptive. In a forthcoming paper we use the commitment level as one of the ingredients of the damage functions of climate policy decision problems, enabling an exploration of trade-offs between long-term and short-term effects. While we do not advocate for prioritizing the first ones over the later, we hope that our metric can contribute to better informed and more sustainable climate decisions.

\begin{appendices}
	
	\section{Extended Data}\label{sec:extended}
	\begin{figure}[h!]
		\centering
		\includegraphics[width=0.8\linewidth]{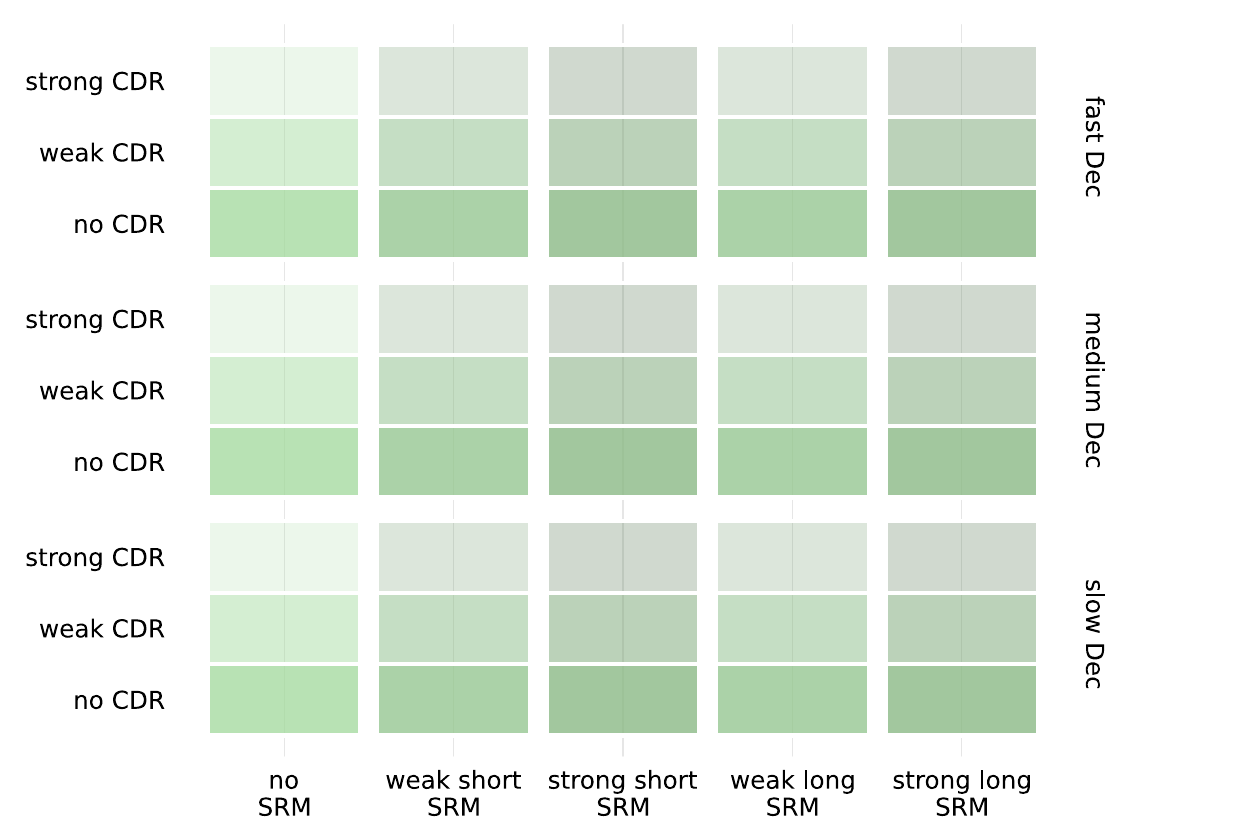}
		\caption{The 45 long-term scenarios considered in the lost options commitment assessment decomposed into different groups. Each column corresponds to a different solar radiation modification option. Rows are divided into groups of three, each of these groups corresponds to different decarbonisation options (right labels). Within each decarbonisation group, the three rows represent different carbon dioxide removal options (left labels).}
		\label{fig:all_options}
	\end{figure}
	\begin{figure}[h!]
		\centering
		\includegraphics[width=\linewidth]{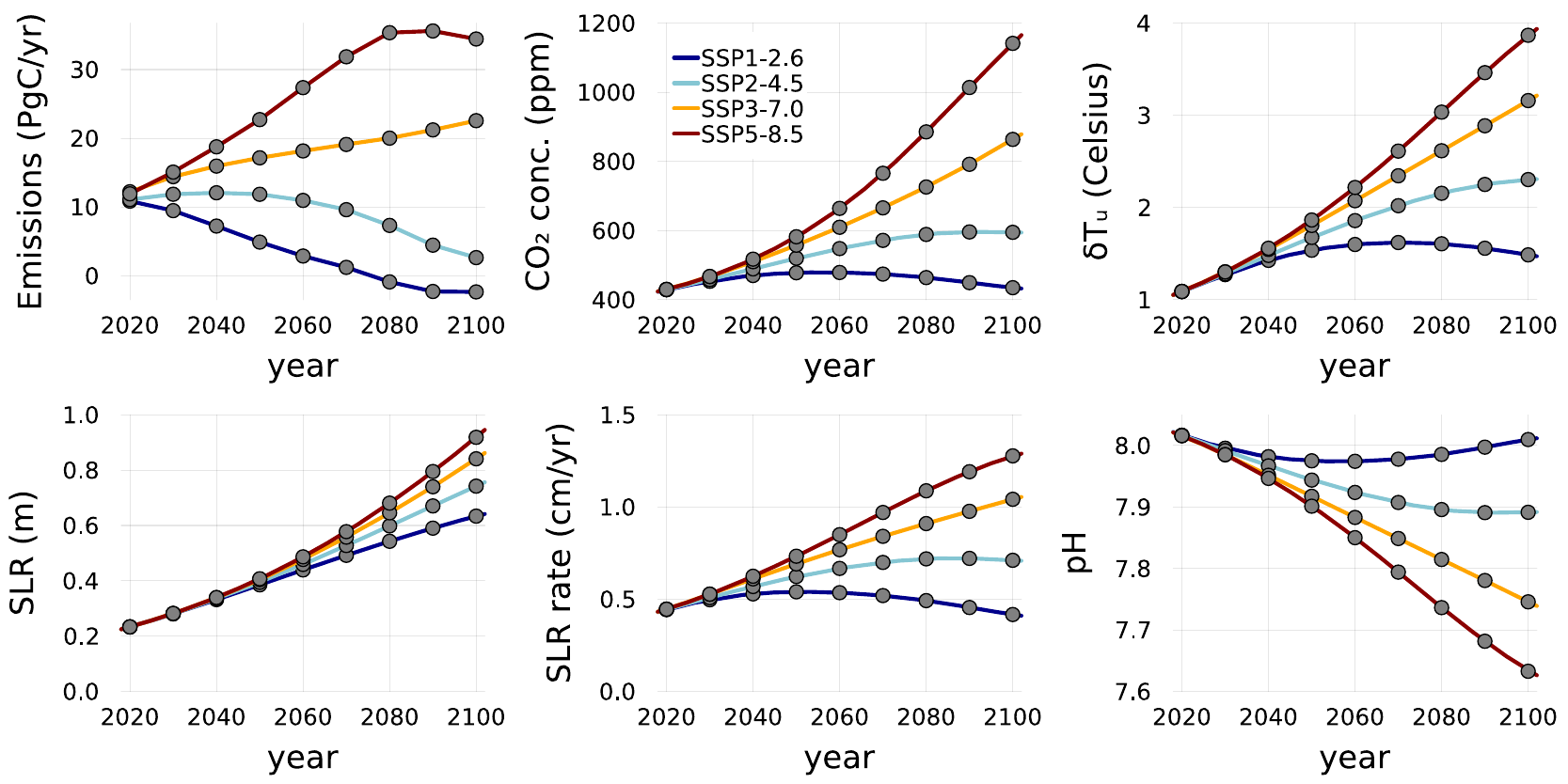}
		\caption{States on which lost options commitment assessment is performed, in grey. Each plot refers to a different quantity characterising the state. From left to right, top to bottom: emission rate, atmospheric CO$_2$ concentration, mean temperature anomaly, sea level rise, sea level rise rate, upper ocean pH.}
		\label{fig:all_states}
	\end{figure}
	\begin{figure}[h!]
		\centering
		\includegraphics[width=\linewidth]{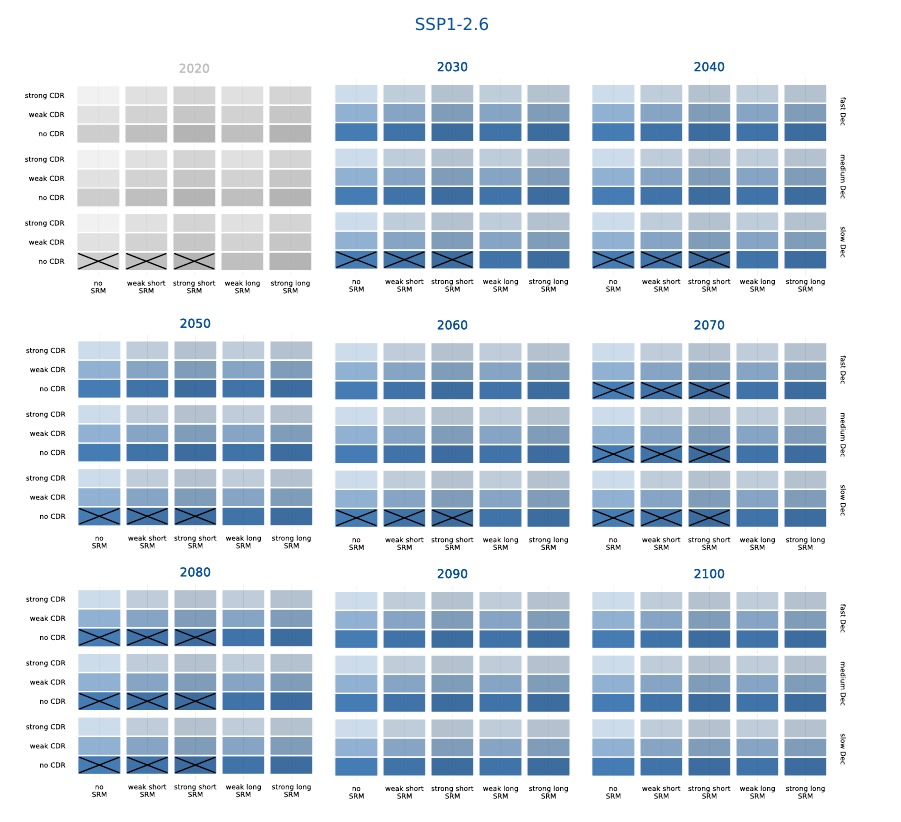}
		\caption{A window of opportunity along SSP1-2.6. The space of considered long-term scenarios is shown for the states assessed. Options that transgress the target are crossed out.}
		\label{fig:SSP1}
	\end{figure}
	\begin{figure}[h!]
		\centering
		\includegraphics[width=\linewidth]{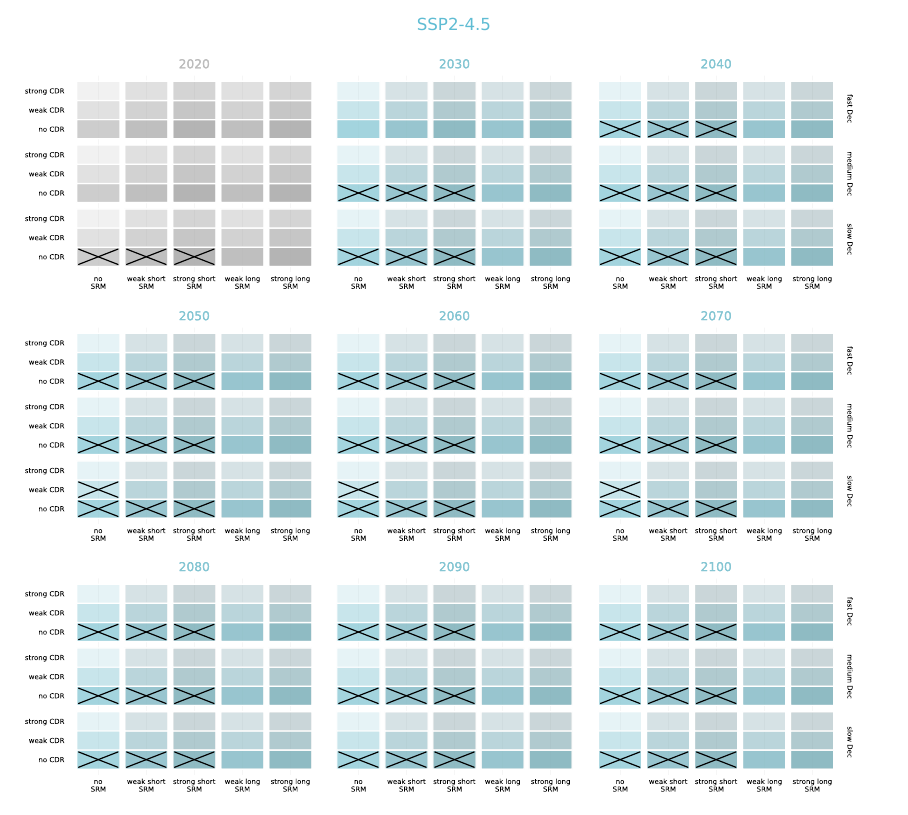}
		\caption{A narrowing window of opportunity along SSP2-4.5. The space of considered long-term scenarios is shown for the states assessed. Options that transgress the target are crossed out.}
		\label{fig:SSP2}
	\end{figure}
	\begin{figure}[h!]
		\centering
		\includegraphics[width=\linewidth]{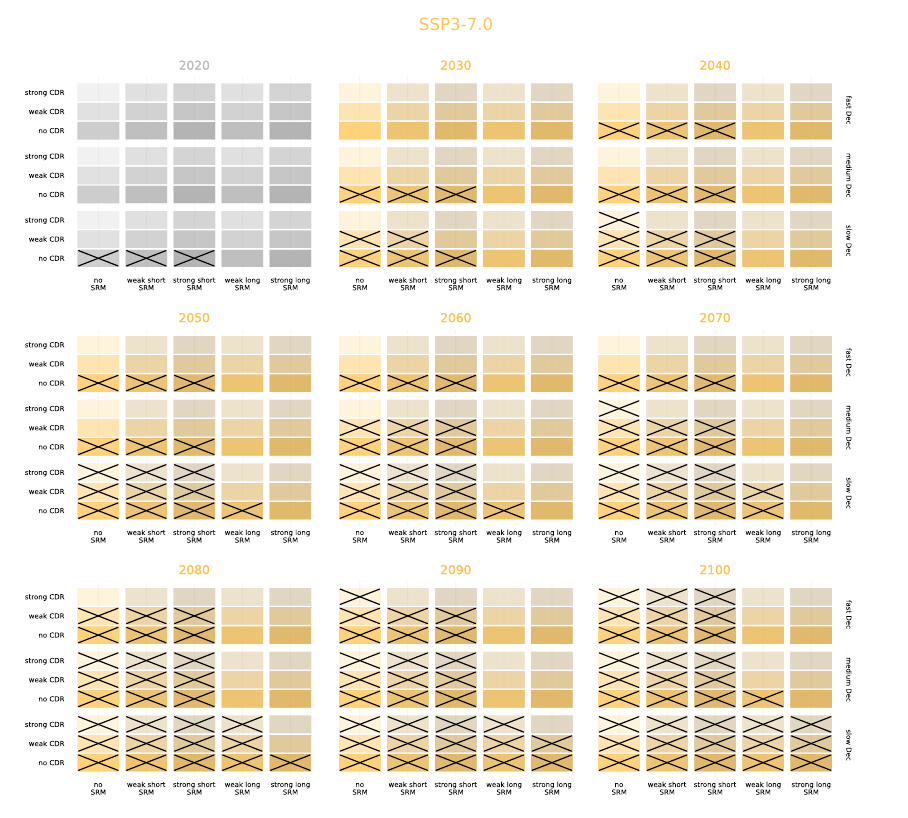}
		\caption{A narrowing window of opportunity along SSP3-7.0. The space of considered long-term scenarios is shown for the states assessed. Options that transgress the target are crossed out.}
		\label{fig:SSP3}
	\end{figure}
	\begin{figure}[h!]
		\centering
		\includegraphics[width=\linewidth]{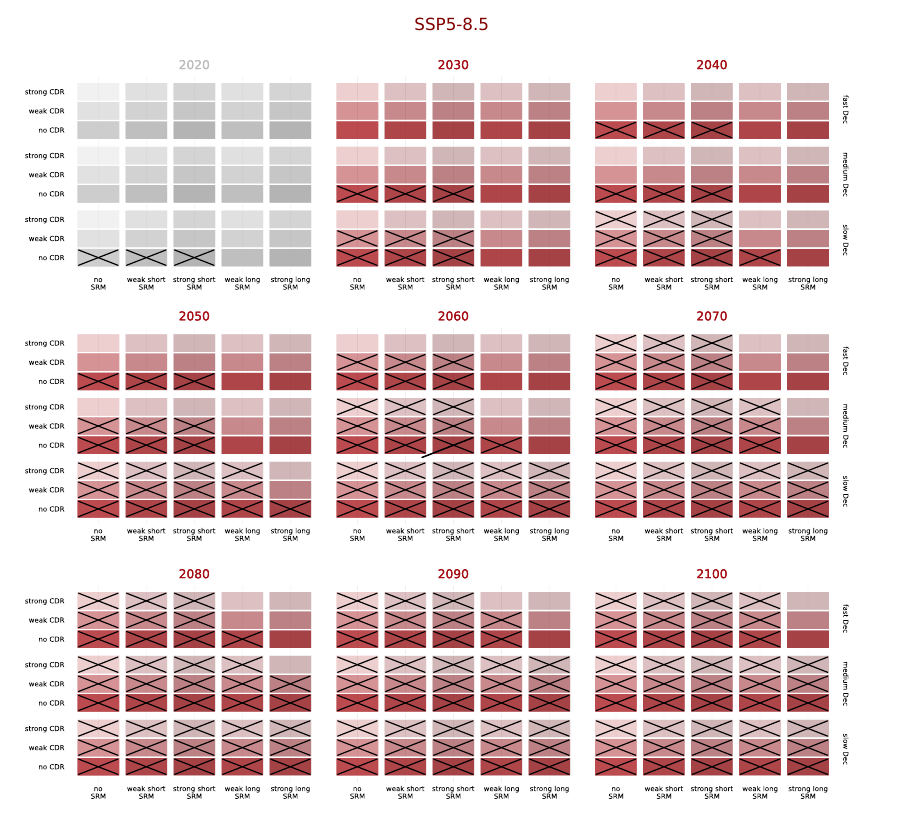}
		\caption{A narrowing window of opportunity along SSP5-8.5. The space of considered long-term scenarios is shown for the states assessed. Options that transgress the target are crossed out.}
		\label{fig:SSP5}
	\end{figure}
	
	\section{Methods}\label{sec:methods}
	To perform the lost options commitment assessment we have relied on the SURFER model \cite{martinez_montero_surfer_2022}. This model has two forcing sources, one corresponding to CO$_2$ emissions ($E$), and another one corresponding to SRM SO$_2$ injections ($I$). The different long-term scenarios present in the lost options commitment assessment correspond to different definitions of these emissions and injections over the assessment period.
	
	The 45 long-term scenarios are constructed as a Cartesian product of three sets of options corresponding to different decarbonisation rates, solar radiation modification intensities and durations, and carbon dioxide removal rates.
	
	Three constant decarbonisation rates are considered: $-0.08,$ $-0.25$ and $-0.75$~GtC/yr$^2$. When applied to a given state with some emission rate, emissions linearly decrease at the specified rate until zero emissions are reached. When applied to a state with zero or negative emissions the three decarbonisation options are degenerate and have no effect.
	
	Five solar radiation modification options are considered: \emph{none}, \emph{weak short}, \emph{strong short}, \emph{weak long}, \emph{strong long}. SRM is used to counter-act all possible global warming and it is achieved through stratospheric aerosol injections. The weak and strong intensities correspond to maximum injection rates of 20 and 40~MtS/yr. The weak intensity corresponds roughly to the injection rates required to reduce the radiative forcing of a high emission scenario like SSP5-8.5 to that of SSP2-4.5 by year 2100 \cite{Visioni2021}, the strong intensity is twice the weak rate. SRM is done ``smartly'' such that just the necessary amount of aerosols are injected to return to pre-industrial temperatures. In SURFER, the precise injection rate needed to flow towards a $\delta T_\text{target}$ is
	\begin{equation}
		I_\text{needed} = \beta_{SO2}\left(-\log\left(\frac{F_{\text{CO}_2}(M_A) - \beta\,\delta T_\text{target}+\gamma(\delta T_D - \delta T_\text{target})}{\alpha_{SO2}}\right)\right)^{-1/\gamma_{SO2}}.\label{eq:Ineeded}
	\end{equation}
	where $M_A$ is the carbon mass in the atmosphere, $F_{\text{CO}_2}$ is the CO$_2$ radiative forcing and $\delta T_D$ is the temperature anomaly of the deeper ocean layer. This equation was obtained by solving
	\begin{equation}
		\frac{d \delta T_{U}}{dt} = 0
	\end{equation}
	for $I = I_\text{needed}$ with $\delta T_U = \delta T_\text{target}$ in Eq.~(29) of \cite{martinez_montero_surfer_2022}, where $\delta T_U$ is the temperature anomaly of the upper ocean layer which is taken as the global mean temperature anomaly. We set $\delta T_\text{target} = 0^{\circ}$C.
	The injection rate used is then
	\begin{equation}
		I_\text{used} = \text{min}\,\{\text{SRM}_\text{limit}, I_\text{needed}\}
	\end{equation}
	where $\text{SRM}_\text{limit}$ is 20 or 40 MtS/yr depending on the long-term scenario. We consider two durations for injection deployment, \emph{short} and \emph{long}. We define \emph{short} as the time it takes to achieve full decarbonisation and \emph{long} to the whole commitment assessment timescale. In the sensitivity analysis presented in the supplementary material we consider other definitions of \emph{short} which are longer than the decarbonisation period.
	
	Three constant atmospheric carbon dioxide removal rates are considered: 0, -1, -2 GtC/yr. CDR takes place since the beginning of the assessment. If decarbonisation has not finished for the particular state there is an immediate reduction in net emissions. CDR continues at the specified constant rate after decarbonisation has ended and until pre-industrial atmospheric CO$_2$ concentration is reached. The CDR rate is then reduced so that it just removes the carbon that flows into the atmosphere from land and ocean, keeping atmospheric concentrations constant at pre-industrial levels until all reservoirs return to pre-industrial conditions. The reduced removal rate is obtained by setting $M_A = M_A(t_{PI})$ in Eq.~(27a) of \cite{martinez_montero_surfer_2022} and solving for the emissions in
	\begin{equation}
		\frac{d M_A}{dt} = 0,
	\end{equation}
	which gives
	\begin{equation}
		\begin{array}{r c c l}
			E_\text{CDR reduced} &=&& k_{A\to U}\left(M_A(t_{PI}) - \frac{m_A}{W_U K_0} B(M_U) M_U\right) \\[1em] &&-& k_{A\to L}\left(M_L - M_L(t_{PI})\right).\label{eq:CDRneeded}
		\end{array}
	\end{equation}
	For details on all the terms in equations \eqref{eq:Ineeded} and \eqref{eq:CDRneeded}, parameter values and pre-industrial conditions we refer the reader to \cite{martinez_montero_surfer_2022}.
	\\\\
	For any given state, the assessment is done by considering 45 different evolutions. Each evolution obtained by evolving the particular state forced by a different long-term scenario, which specifies the emission and injection forcings, for 2000 years. This results in 45 trajectories, each corresponding to a different long-term scenario. Each trajectory is then inspected against the chosen target, in the presented example, whether sea level rise is higher than 3 meters above pre-industrial values within those 2000 years. We integrate SURFER's differential equations in Julia using the package \texttt{DifferentialEquations.jl} with the integration method \texttt{Rosenbrock23()}, \texttt{abstol=1e-12} and \texttt{reltol=1e-3}.
	
\section{Supplementary Information}\label{sec:suplementary}
The lost options commitment assessment needs the specification of three ingredients:
\begin{itemize}
	\item time horizon
	\item target
	\item long-term scenarios (space of options)
\end{itemize}
The results of the assessment will then, by construction, depend on these ingredients, which define the nature of the assessment itself. However, for the assessment to be useful and reasonable, robustness to small changes in criteria is required. In what follows, we perform three sensitivity analyses for the example discussed in the main text and discuss how the presence of tipping points in
the climate system might impact our metric.

\begin{figure}
	\centering
	\includegraphics[width=\linewidth]{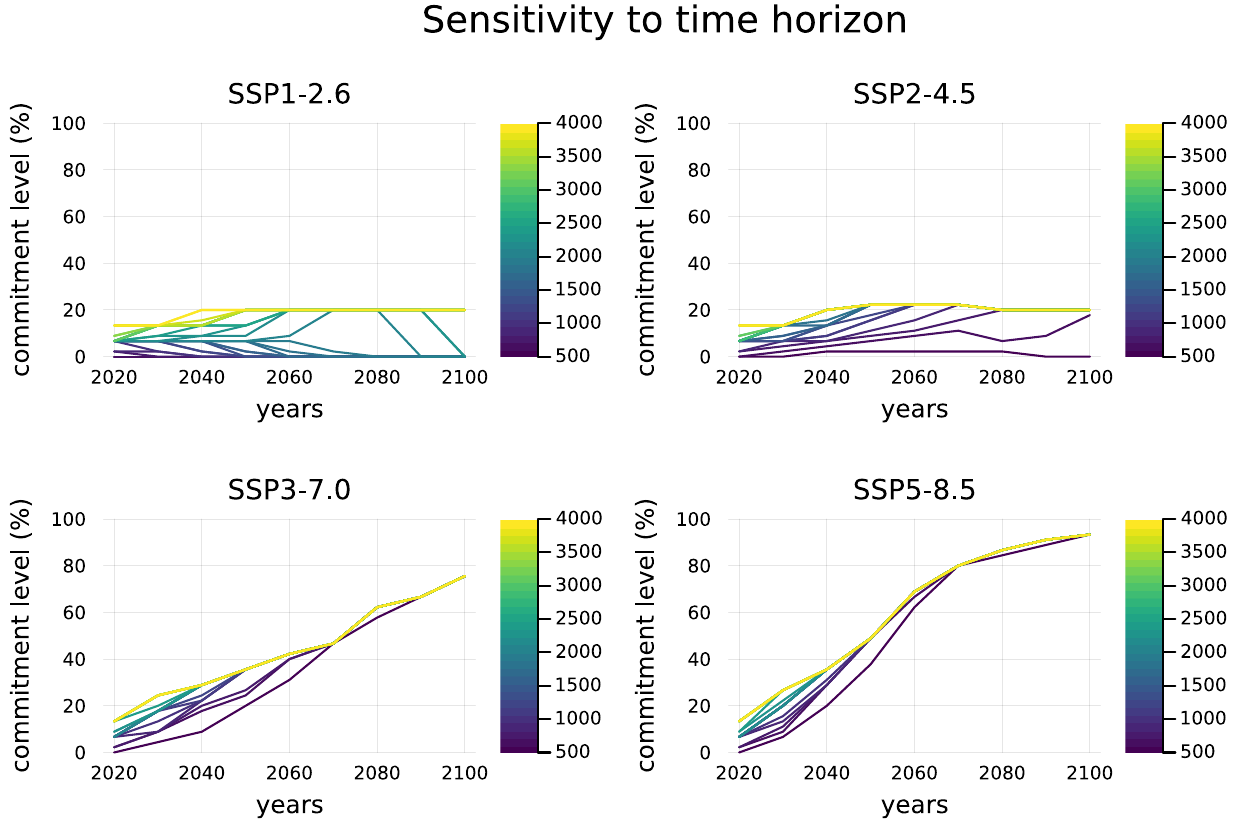}
	\caption{Sensitivity of commitment level to time horizon. The target is fixed as sea level rise higher than 3 meters within the time horizon. 30 equally spaced time horizons have been considered in the range between 500 to 4000 years.}
	\label{fig:horizon_sens}
\end{figure}

\paragraph{Sensitivity to time horizon}
First we explore the sensitivity of the commitment level to the time horizon. We repeat the main text example with time horizons from 500 to 4000 years, see Fig.~\ref{fig:horizon_sens}.
There we see several things:
\begin{enumerate}
	\item Longer horizons result in higher commitment levels. This is because sea level rise is a slow process. If the horizon is too short, a given option might respect the specified target, while when extending the horizon that same option might lead to target transgression.
	\item However, for horizons of 2500 to 4000 years, the metric converges because:
	\begin{enumerate}
		\item Only some of the long-term scenarios exhibit temperatures that correspond to committed sea level rise higher than the target. Long-term scenarios that do not exhibit high temperatures for long enough do not lead to high sea level rise, no matter how long the time horizon.
		\item A large fraction of the committed sea level rise happens within the next 3000 years, after that time, all options that would eventually transgress the target of 3m have already done so. 
	\end{enumerate}
	\item The presence of ice sheet tipping points in SURFER is not apparent in these particular results. This is reasonable since the target of 3m is lower than the sea level rise potential of Greenland ice sheet, which is  around 7m, and hence does not differentiate between tipping and non-tipping options.
\end{enumerate}

\begin{figure}
	\centering
	\includegraphics[width=\linewidth]{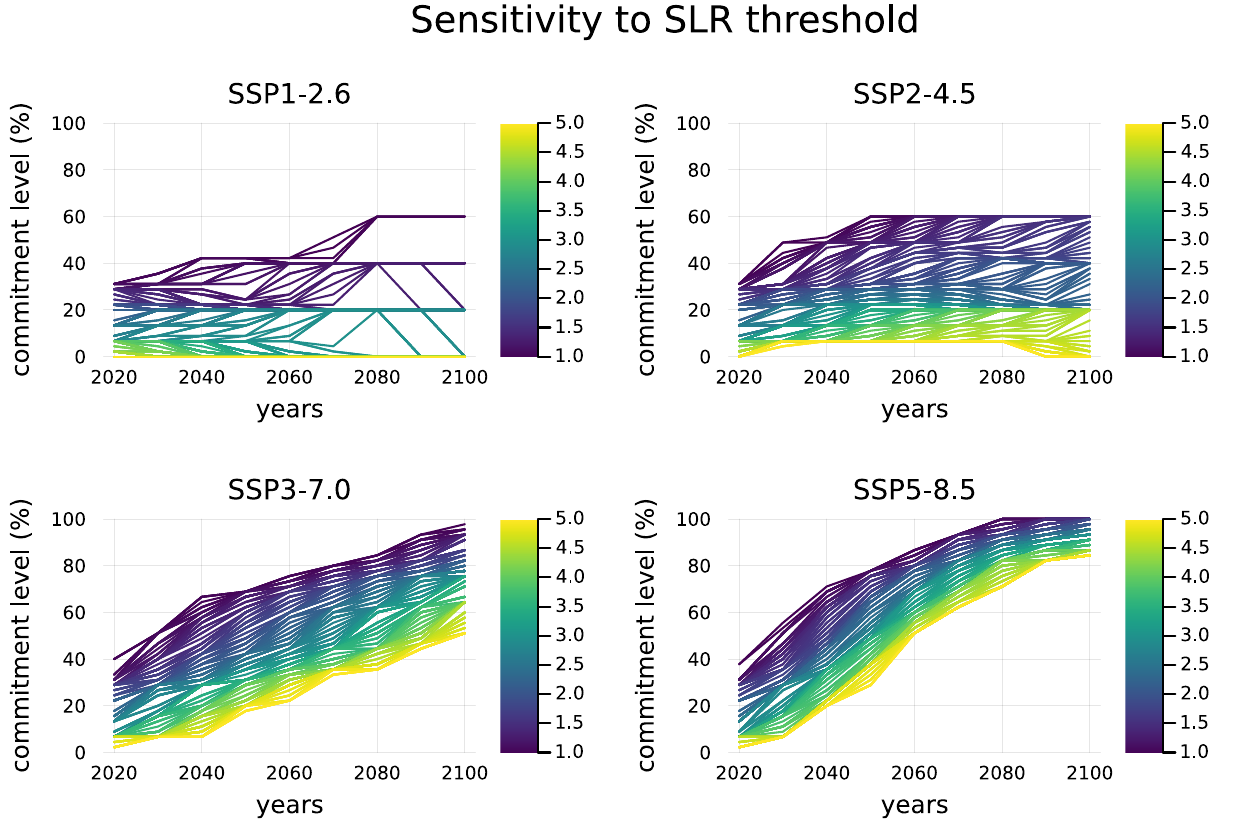}
	\caption{Sensitivity of commitment level to sea level rise target. 400 thresholds in the range between 1 to 5 meters have been considered. Time horizon is fixed at 2000 years.}
	\label{fig:threshold_sens}
\end{figure}

\paragraph{Sensitivity to targets}
Second, we analyse the sensitivity to the target, specifically, to the threshold chosen for sea level rise. We repeat the  example in the main text, but with varying thresholds from 1 to 5 meters, see Fig.~\ref{fig:threshold_sens}.
There we see that:
\begin{enumerate}
	\item Lower SLR thresholds lead to higher commitment levels.
	\item Gradual changes in SLR threshold lead to gradual changes in commitment levels. States along SSP1-2.6 after 2070 are an exception to this.
	\item For states after 2070 along of SSP1-2.6 certain ranges of targets exhibit degenerate commitment levels. In particular, by 2100, all considered thresholds (400) collapse into 4 groups with commitment levels: 0\%, 20\%, 40\% and 60\%. This can be understood due the big degeneracy in the long-term scenarios due to the fact that decarbonisation has finished:
	\begin{enumerate}
		\item The three decarbonisation speeds are degenerate in this case.
		\item The \emph{short SRM} options (defined to happen during decarbonisation) are degenerate with the \emph{no SRM} option.
	\end{enumerate}
\end{enumerate}
\begin{figure}
	\centering
	\includegraphics[width=\linewidth]{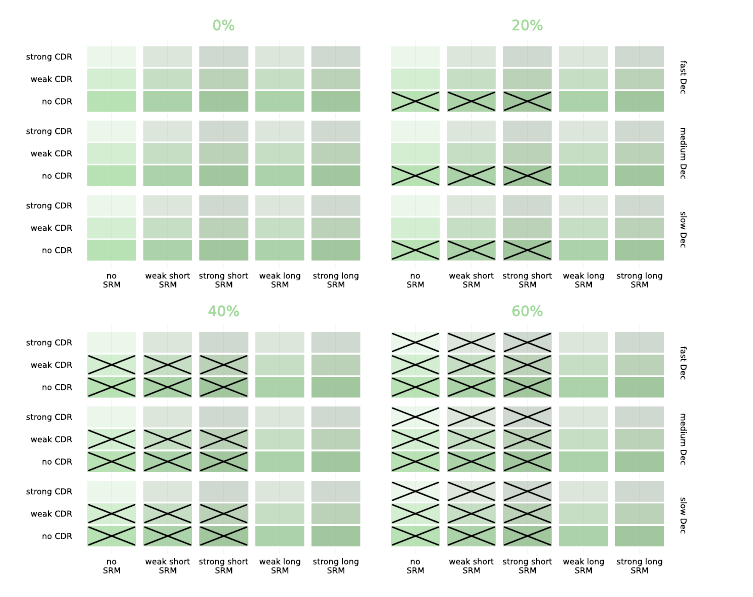}
	\caption{Available options in the 4 commitment level groups resulting from the target sensitivity experiment for SSP1-2.6 in 2100. The percentage in the title indicates the commitment level of the given group.}
	\label{fig:SSP1_groups}
\end{figure}
The available options corresponding to these four groups can be found in Fig.~\ref{fig:SSP1_groups}. Focusing on the 60\% commitment level group, which corresponds to relatively low sea level rise thresholds, we see that CDR rates considered in the long-term scenarios are not large enough to stop a sea level rise of ~1 meter in the next 2000 years without relying on SRM for the state corresponding to SSP1-2.6 at 2100.

\paragraph{Sensitivity to long-term scenarios}
In Figs.~\ref{fig:SSP1},~\ref{fig:SSP2},~\ref{fig:SSP3},~\ref{fig:SSP5} we see an almost absolute degeneracy between the long-term scenarios that have \emph{no SRM} and those that have \emph{short SRM}. This suggests that SRM until the end of decarbonisation is too short to have a significant impact - especially in the absence of CDR, as was already shown by \cite{baur_deployment_2023}. In such scenarios, SRM stops shortly after the peak in CO$_2$ concentration. For this reason, we consider different definitions of \emph{short SRM}, and assess the sensitivity of the lost options commitment assessment to these definitions. Specificaly, we set the duration of \emph{short SRM} to be that of decarbonisation plus an extension which ranges from 0 to 300 years. This corresponds to a policy of buying enough time for natural feedbacks to kick in. 
\begin{figure}
	\centering
	\includegraphics[width=\linewidth]{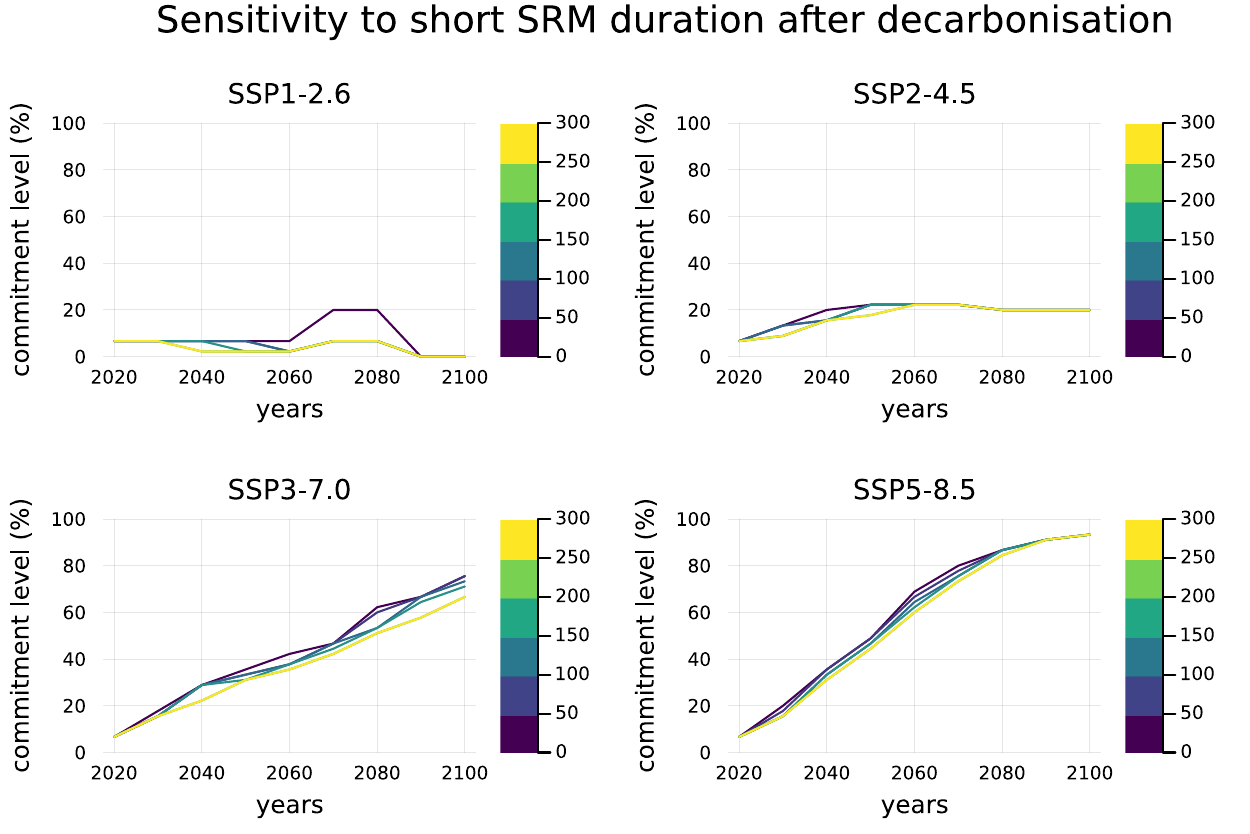}
	\caption{Sensitivity of commitment level to the duration of the \emph{short SRM} options. The color scale corresponds to the years of continued SRM after decarbonisation has finished.}
	\label{fig:sens_shortSRM}
\end{figure}
Figure~\ref{fig:sens_shortSRM} shows that overall, extending the duration of \emph{short} SRM deployement decreases the commitment level, as expected. The effect is relatively small for medium and high emission scenarios (SSP2-4.5, SSP3-7.0 and SSP5-8.5) but is more significant for SSP1-2.6, in 2070 and 2080.
\begin{figure}
	\centering
	\includegraphics[width=0.9\linewidth]{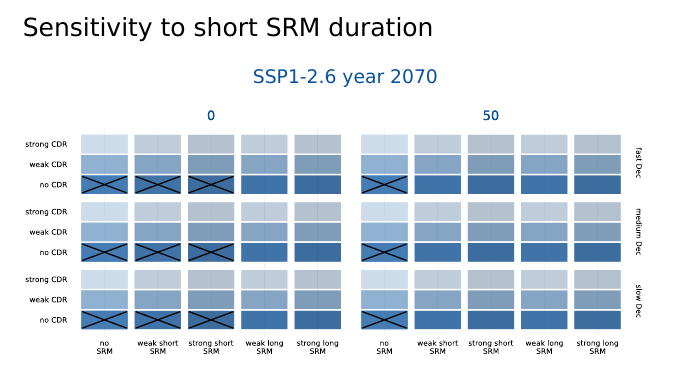}
	\caption{Sensitivity of available options to \emph{short SRM} duration for SSP1-2.6 in year 2070. The numbers indicate the years of continued SRM after decarbonisation has finished. Only the columns corresponding to \emph{short SRM} are subject to change.}
	\label{fig:sens_shortSRM_SSP1}
\end{figure}
\begin{figure}
	\centering
	\includegraphics[width=\linewidth]{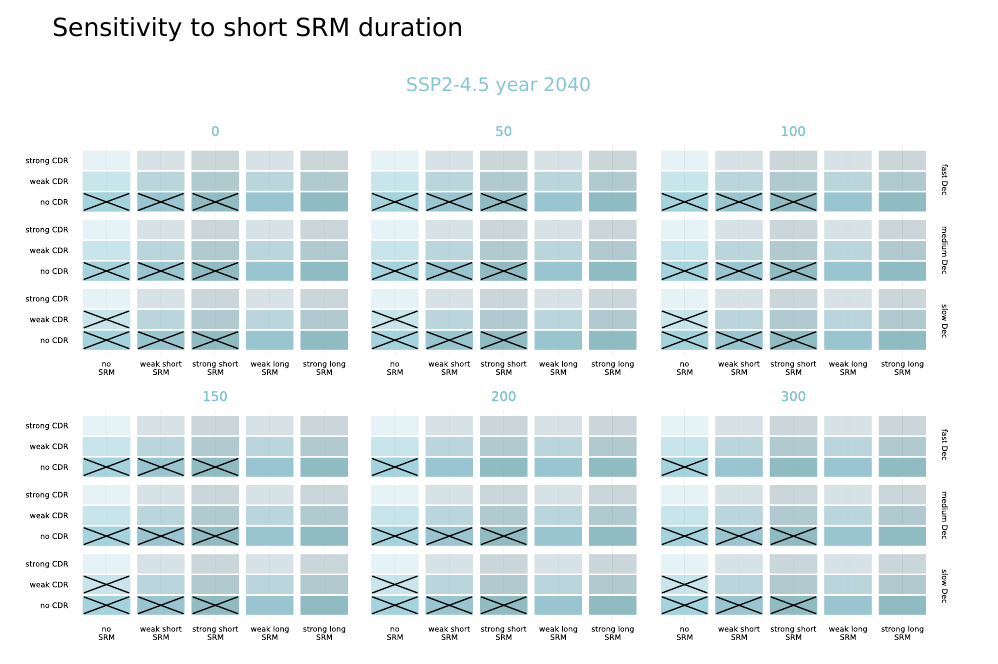}
	\caption{Sensitivity of available options to \emph{short SRM} duration for SSP2-4.5 in year 2040. The numbers indicate the years of continued SRM after decarbonisation has finished. Only the columns corresponding to \emph{short SRM} are subject to change.}
	\label{fig:sens_shortSRM_SSP2}
\end{figure}
\begin{figure}
	\centering
	\includegraphics[width=\linewidth]{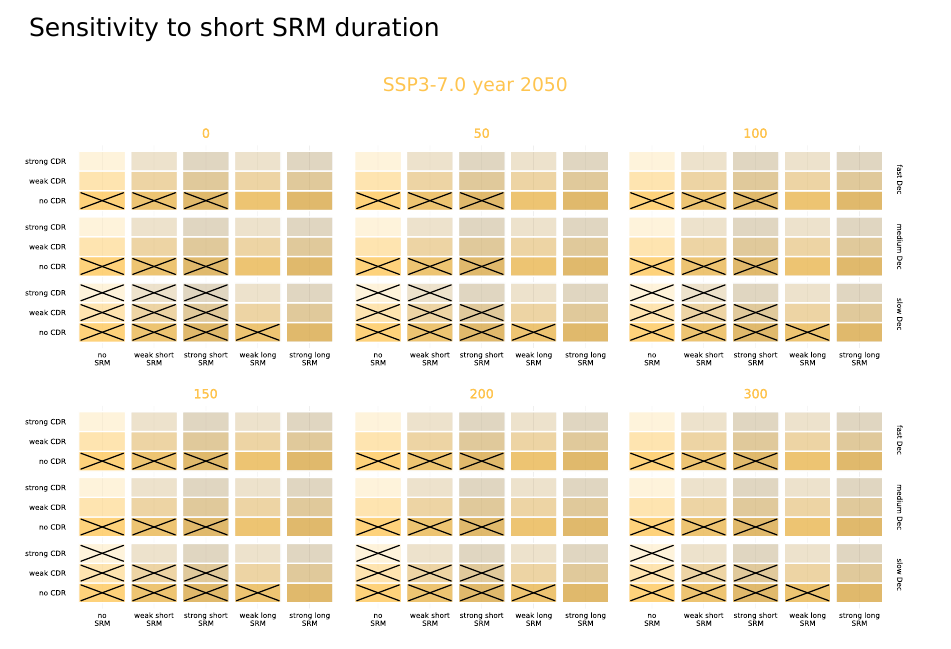}
	\caption{Sensitivity of available options to \emph{short SRM} duration for SSP3-7.0 in year 2050. The numbers indicate the years of continued SRM after decarbonisation has finished. Only the columns corresponding to \emph{short SRM} are subject to change.}
	\label{fig:sens_shortSRM_SSP3}
\end{figure}
\begin{figure}
	\centering
	\includegraphics[width=\linewidth]{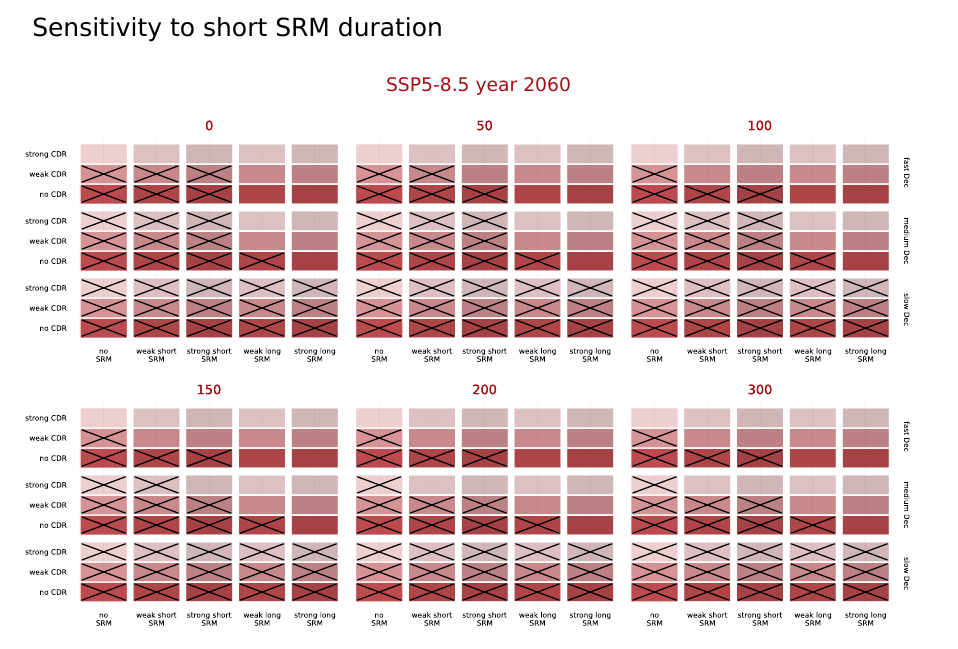}
	\caption{Sensitivity of available options to \emph{short SRM} duration for SSP5-8.5 in year 2060. The numbers indicate the years of continued SRM after decarbonisation has finished. Only the columns corresponding to \emph{short SRM} are subject to change.}
	\label{fig:sens_shortSRM_SSP5}
\end{figure}
Figures~\ref{fig:sens_shortSRM_SSP1}, \ref{fig:sens_shortSRM_SSP2}, \ref{fig:sens_shortSRM_SSP3} and \ref{fig:sens_shortSRM_SSP5} show the space of options for different \emph{short SRM} durations for a particular state. Extending the duration of \emph{short SRM} has the effect of buying time and making some of the transgressing options viable ones. Depending on the emission rate associated to the state being assessed, the liberated options are different: 
\begin{itemize}
	\item For SSP1-2.6 in year 2070 decarbonisation has almost finished, see Fig.~\ref{fig:all_states}. Extending the \emph{short SRM} period frees up the options with \emph{no CDR, short SRM} independently of the decarbonisation rate, see Fig.~\ref{fig:sens_shortSRM_SSP1}, i.e., it buys time for natural sinks to kick in.
	\item For SSP2-4.5 in year 2040 emissions are high, see Fig.~\ref{fig:all_states}. Extending the \emph{short SRM} period frees up the options with \emph{fast Dec, no CDR, short SRM}, see Fig.~\ref{fig:sens_shortSRM_SSP2}. In this case SRM also buys time for natural sinks to kick in. Notice however that it does not buy enough time for \emph{medium Dec} and \emph{slow Dec} to become viable.
	\item For SSP3-7.0 in year 2050, we see in Fig.~\ref{fig:sens_shortSRM_SSP3} that extending the \emph{short SRM} period enables the options with \emph{slow Dec, strong CDR, short SRM}. In this case a longer \emph{short SRM} period allows for a slower decabonisation with strong CDR deployment to be a viable option.
	\item Finally, for SSP5-8.5 in year 2060, Fig.~\ref{fig:sens_shortSRM_SSP5}, extending the \emph{short SRM} period enables the options with \emph{fast Dec, medium CDR, short SRM} and extending it further, the options  with \emph{medium Dec, strong CDR, short SRM}.
\end{itemize}

\paragraph{Ice sheet tipping points and lost options commitment}
We have stated that for the metric to be useful and reasonable some robustness to small criteria variations is required. We see two caveats to this argument. First, lost options commitment could be sensitive to very small variations in the target if several of the long-term scenarios produce similar or identical trajectories. This is what we observed for states after 2070 along of SSP1-2.6 in Fig.~\ref{fig:threshold_sens}. Second, lost options commitment could be sensitive to very small variations in the time horizon if abrupt changes are observed in the target variable. This could typically happen in the presence of fast climate tipping elements. In this particular study, however, the only tipping elements considered are the ice sheets, which exhibit slow dynamics, and so we don't expect our metric to be very sensitive to small changes in the time horizon.

\begin{figure}[h]
	\centering
	\includegraphics[width=0.7\linewidth]{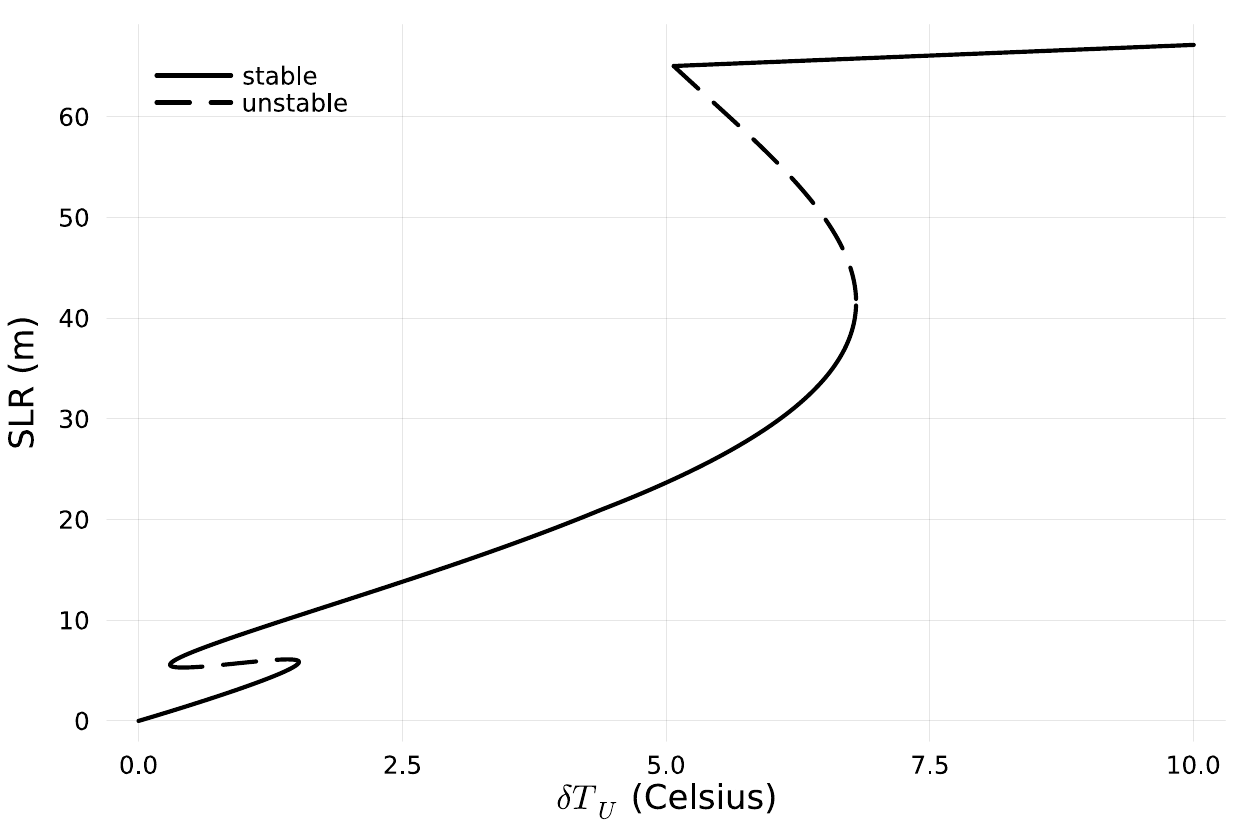}
	\caption{Sea level rise bifurcation diagram for the SURFER model.}
	\label{fig:bif_SLR_T}
\end{figure}

\begin{figure}[h]
	\centering
	\includegraphics[width=\linewidth]{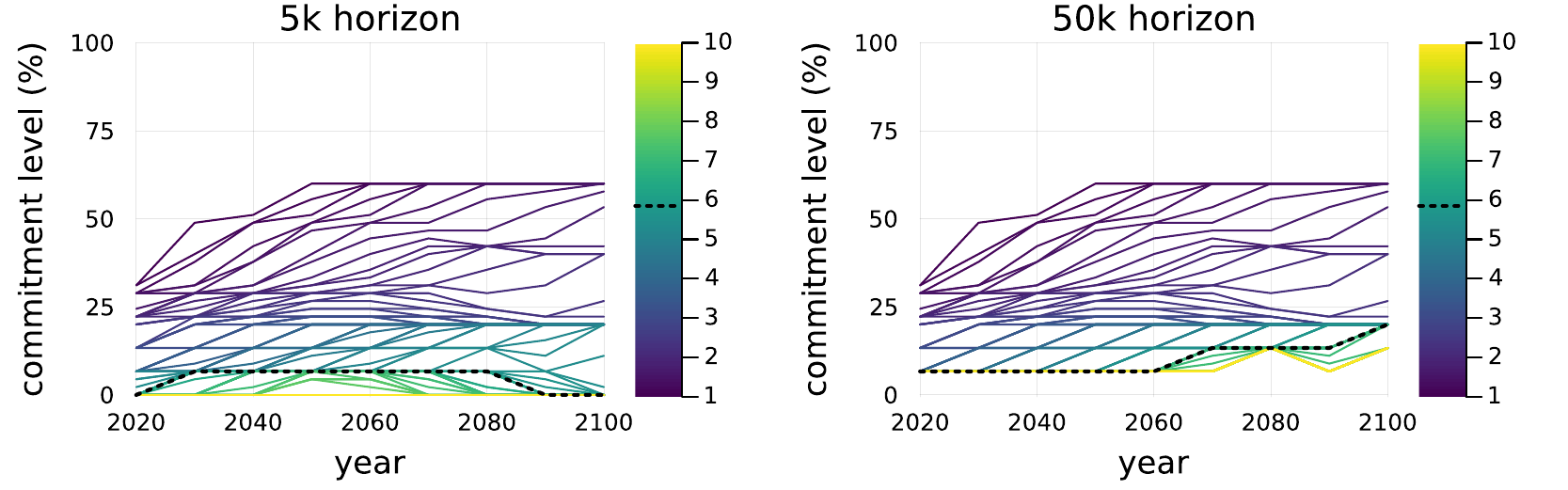}
	\caption{Sensitivity to target for SSP2-4.5 for two different time horizons. Dashed black line corresponds to the SLR target coinciding with the SLR tipping point.}
	\label{fig:sens_target_tipping}
\end{figure}

In the presented sensitivity analyses of the previous sections, the ice sheet tipping points present in the SURFER model did not have an effect on the lost options commitment assessment. These analyses, however, were by design incapable of showcasing such an effect because of two reasons:
\begin{enumerate}
	\item We did not explore a big enough sea level rise target.
	\item We did not explore big enough time horizons.
\end{enumerate}
In the default setup of the SURFER model, Greenland's ice sheet tipping point is at a global mean temperature anomaly of 1.52$^\circ$C. The committed sea level rise from all sea level rise contributors at that temperature corresponds to $\approx 5.9$m for the state with ice on Greenland and to more than 10m for the state with ice free Greenland, see Fig.~\ref{fig:bif_SLR_T}.
In order to see whether Greenland's ice sheet tipping point has an effect on the commitment assessment, the sea level rise target should be set to 5.9m or larger. We repeated the target sensitivity experiment for sea level rise targets from 1m to 10m for states along SSP2-4.5 considering a time horizon of 5kyr and a time horizon of 50kyr, see Fig.~\ref{fig:sens_target_tipping}. 
For both considered horizons we see that gradual changes in the target lead to gradual differences in the commitment level. There we see increases in the commitment level of consecutive states for different climate targets. For targets smaller than $\sim 5.9$m those increases are not necessarily related to Greenland's tipping point. We also see that both horizons lead to identical commitment levels when the targets are small enough (from 1m to 5m approx.) and that they differ for targets slightly smaller than $\sim 5.9$m to 10m.

\begin{figure}[h]
	\centering
	\includegraphics[width=\linewidth]{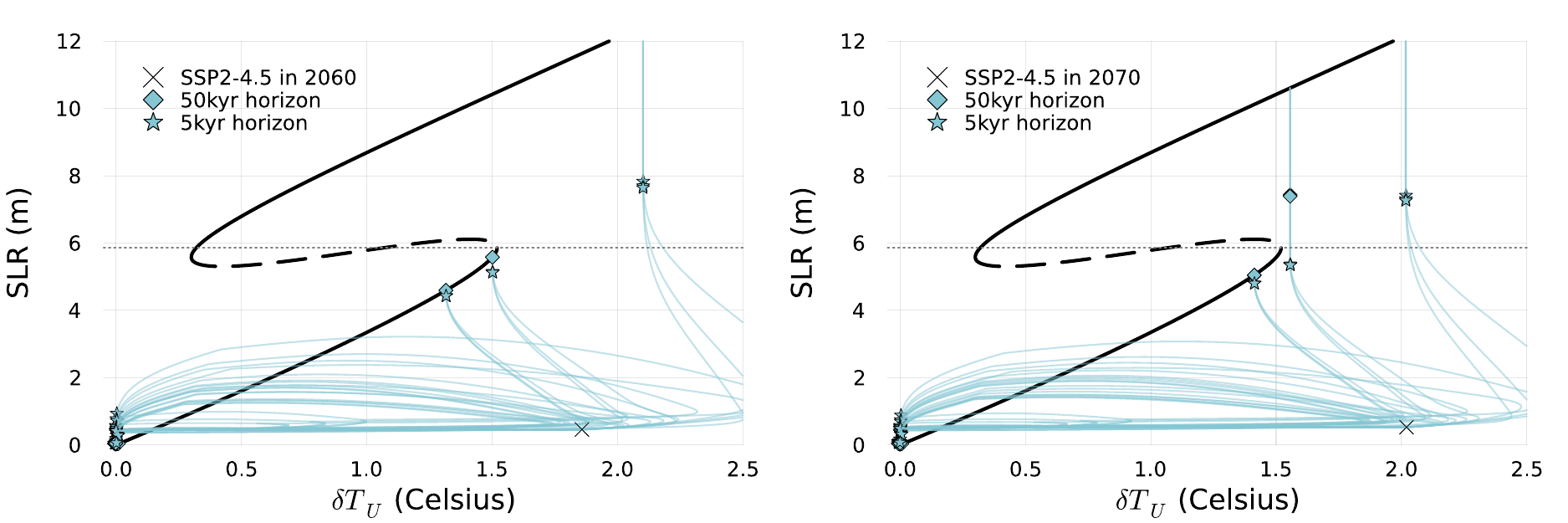}
	\caption{Trajectories corresponding to the different long-term scenarios starting from SSP2-4.5 at 2060 (left) and 2070 (right), in light blue. Thick black likes correspond to the SLR steady states of the SURFER model. The x marker indicates the initial state, the star and diamond markers the position in phase space after 5kyr and 50kyr respectively.}
	\label{fig:phase_SSP2}
\end{figure}

For the 50k horizon with target $\sim 5.9$m there are increases in the commitment level for the state from 2060 to 2070 and 2090 to 2100. These increases in commitment level happen because a group of options that end on the lower SLR branch of the bifurcation diagram for the lower committed state (e.g., SSP2-4.5 at 2060), end on the higher SLR branch for the consecutive state (in this case, SSP2-4.5 at 2070), this can be seen in Fig.~\ref{fig:phase_SSP2}. When this happens the trajectories associated with those lost options pass very close to the ice sheet bifurcation point in phase space. The dynamics close to a bifurcation point tend to be very slow, and a time horizon of 5kyr is not enough to appreciate the committed SLR. We put time markers in  Fig.~\ref{fig:phase_SSP2} to make this point and help understand the sensitivity shown on Fig.~\ref{fig:sens_target_tipping}.
We conclude that very large time horizons, together with appropriate targets, are needed for tipping behaviour to be apparent in the lost options commitment assessment. This is because the already slow ice sheet dynamics slow down even more close to bifurcation points. We have shown that even when considering very long horizons, the impact of ice sheet tipping points on lost options commitment assessment isn't big. The increases seen in commitment level in Fig.~\ref{fig:sens_target_tipping} for bigger targets and 50k horizon is of the same order than the ones corresponding to other targets and timescales. Additionally, SURFER v.2.0 does not include sediment dynamics related to carbonate dissolution and silicate weathering, which determine the fate of fossil fuels at time scales beyond several millennia. Those additional processes act as  CO$_2$ sinks and have the potential of lowering the commitment level on assessments with such long horizons, making it even harder for ice sheet tipping points to show up as commitment level tipping points in such lost options commitment assessments. 
\end{appendices}

\section{Competing interests}
No competing interest is declared.

%\section{Author contributions statement}

\section{Acknowledgments}
This project is TiPES contribution No. 237: this project has received funding from the European Union’s Horizon 2020 research and innovation programme under grant agreement no. 820970. Michel Crucifix is funded as Research Director by the Belgian National Fund of Scientific Research. Victor Couplet is funded as Research Fellow by the Belgian National Fund of Scientific Research (F.S.R.-FNRS). C.W. is supported by the Dutch government under the Sectorplan science and technology.

\bibliography{model.bib}
\bibliographystyle{unsrt}

\end{document}